\documentclass[aps,pra,reprint,superscriptaddress]{revtex4-1}

\usepackage{graphicx}
\usepackage{amsmath}
\usepackage{amsthm}
\usepackage{dsfont}

\begin{document}

\title{Disordered Haldane-Shastry model}

\author{Shriya Pai}
\thanks{These authors contributed equally to this work. Email of corresponding author: srivatsa@pks.mpg.de}
\affiliation{Max-Planck-Institut f\"{u}r Physik komplexer Systeme, D-01187 Dresden, Germany}
\affiliation{Department of Physics, Indian Institute of Science, Bangalore 560012, India}
\affiliation{Department of Physics, University of Colorado, Boulder, CO 80309, USA}

\author{N. S. Srivatsa}
\thanks{These authors contributed equally to this work. Email of corresponding author: srivatsa@pks.mpg.de}
\affiliation{Max-Planck-Institut f\"{u}r Physik komplexer Systeme, D-01187 Dresden, Germany}

\author{Anne E. B. Nielsen}
\altaffiliation{On leave from Department of Physics and Astronomy, Aarhus University, DK-8000 Aarhus C, Denmark}
\affiliation{Max-Planck-Institut f\"{u}r Physik komplexer Systeme, D-01187 Dresden, Germany}

\begin{abstract}
The Haldane-Shastry model is one of the most studied interacting spin systems. The Yangian symmetry makes it exactly solvable, and the model has semionic excitations. We introduce disorder into the Haldane-Shastry model by allowing the spins to sit at random positions on the unit circle and study the properties of the eigenstates. At weak disorder, the spectrum is similar to the spectrum of the clean Haldane-Shastry model. At strong disorder, the long-range interactions in the model do not decay as a simple power law. The eigenstates in the middle of the spectrum follow a volume law, but the coefficient is small, and the entropy is hence much less than for an ergodic system. In addition, the energy level spacing statistics is neither Poissonian nor of the Wigner-Dyson type. The behavior at strong disorder hence serves as an example of a non-ergodic phase, which is not of the many-body localized kind, in a model with long-range interactions and SU(2) symmetry.
\end{abstract}

\maketitle

\section{Introduction}

The typical behavior of interacting quantum many-body systems is to display ergodic physics. This is encoded in the eigenstate thermalization hypothesis, which says that local observables computed for an individual eigenstate of the Hamiltonian is described by the thermal microcanonical ensemble \cite{Deu,Sred,Srednicki1999,Rigol2008,Luca,Tarun}. The eigenstate thermalization hypothesis fails, however, in quantum systems that possess integrability and in systems that many-body localize due to quenched disorder where a notion of emergent integrability appears \cite{lbits2,lbits,Rahul,Imbrie2016}. Many-body localization is a phenomenon where an interacting quantum many-body system fails to thermalize and memory of the initial conditions of the local observables is retained even at long times.

Symmetries play an important role in deciding the fate of the excited states. For instance, the presence of a global discrete symmetry of the Hamiltonian may lead to a glassy phase in the excited states \cite{order,Bard,Parameswaran_2018}. However, disordered many-particle systems that possess non-Abelian continuous symmetries fail to many-body localize as pointed out in Refs.\ \cite{vasseur,potter,SU2,nonab}. It was, e.g., argued that the highly excited states of one-dimensional spin models with SU(2) symmetry cannot have area law entanglement as required for many-body localization. Instead, an investigation of the antiferromagnetic Heisenberg chain with SU(2) symmetry and nearest neighbor exchange interactions of random strengths showed that a different type of non-ergodic phase appears in a broad regime at strong disorder \cite{SU2,nonab}.

Among the spin models in one dimension, the Haldane-Shastry model \cite{HSmodelH,shast} is special for several reasons. The Haldane-Shastry model is a long-ranged Heisenberg antiferromagnet with $1/r^2$ exchange interactions. It possesses a Yangian symmetry that makes it fully integrable \cite{FDM1,FDM2,talstra,Talstra_1995}. All the eigenstates and the energies can be computed exactly. In addition, this model is known to be the conformal gapless fixed point of the spin models within the $\textrm{SU(2)}_1$ WZW class, a Luttinger liquid \cite{HSmodelH,affleck,FDM5}. The excitations can be understood as a gas of free semions \cite{FDM3,Greiter,FDM4}.

In the Haldane-Shastry model, the spins are distributed equidistantly on the unit circle. A generalization of the Haldane-Shastry Hamiltonian, which allows the spins to sit at arbitrary positions on the unit circle, was proposed in Refs.\ \cite{cirac,nielsen2011quantum}. This opens up the possibility to study the effect of disorder by introducing randomness into the chosen positions of the spins. The ground state is still known analytically, and different disorder averaged properties of the ground state have been computed in \cite{cirac,stephan2016full,srivatsa2020many}.

In this paper, we study the effects of disorder on the excited states of the Haldane-Shastry model. The randomness introduced into the positions of the spins on the unit circle leads to randomness in the spin-spin couplings. The strengths of the spin-spin couplings do not follow a simple power law, and the couplings can be large even for spins that are not nearest neighbors.

At weak disorder, the model possesses an approximate Yangian symmetry due to the proximity to the clean model, which is integrable, and the disorder averaged entanglement entropy follows a volume law. At strong disorder, the entanglement entropy also follows a volume law, but the coefficient is much smaller than for weak disorder, and the entanglement is hence much less than for an ergodic system. To further confirm this non-ergodic phase, we also study the level spacing statistics at strong disorder, which turns out to be neither Poissonian nor of the Wigner-Dyson type.

There is currently much interest in finding out how nonlocal terms in a Hamiltonian affect the thermalization properties of a system \cite{yao,singh,Nand,nag}, e.g.\ whether many-body localization is possible when different types of long-range terms are present, as one might expect that long-range terms could increase the amount of entanglement.  As mentioned above, it has been argued that systems with SU(2) symmetry cannot many-body localize, since SU(2) symmetry is incompatible with area law scaling of the entanglement entropy \cite{vasseur,potter}, and it is hence expected that we do not find many-body localization in the studied, nonlocal model. Here, we are interested in the question, whether the non-ergodic, low entanglement entropy behavior found in nearest neighbor models with SU(2) symmetry can also appear in models with long-range terms. Our results provide an example showing that this is indeed the case.

The paper is structured as follows. In Sec.\ \ref{sec:model}, we discuss the Haldane-Shastry model and how we introduce disorder into the system. In Sec.\ \ref{sec:exc}, we study the physics of the highly excited states by probing the entanglement entropy and the level spacing statistics. We find that the system at strong disorder is in a non-ergodic phase, which is not a many-body localized phase. In Sec.\ \ref{sec:last}, we briefly discuss an alternative approach to adding disorder to the model. Sec.\ \ref{sec:conclusion} concludes the paper.

\section{The Haldane-Shastry model and disorder}\label{sec:model}

\begin{figure}
\includegraphics[width=\columnwidth]{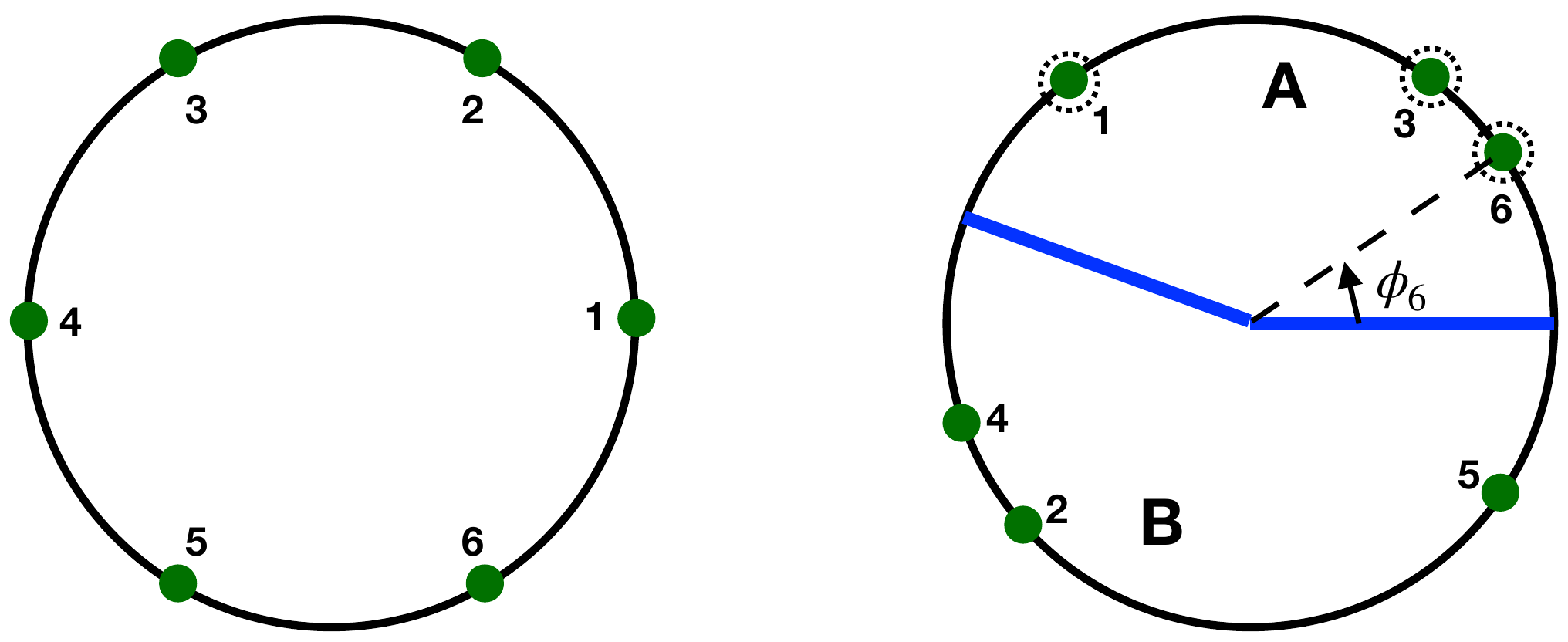}
\caption{Left: Uniform lattice on a circle with $N=6$ spins. Right: Disordered lattice with $N=6$ spins. When we divide the system into two parts to compute the entanglement entropy, we choose part $A$ of the system to be the $N/2$ spins with the smallest angles $\phi_j$ and part $B$ to be the remaining spins.}\label{fig:circle}
\end{figure}

We first recall the generalized Haldane-Shastry model from \cite{nielsen2011quantum,cirac} and discuss how we introduce disorder into the model. Consider a system of $N$ spin-$1/2$ particles sitting on the unit circle as illustrated in Fig.\ \ref{fig:circle}. We denote the two basis states of the $j$th spin as $|s_j\rangle$ with $s_j\in\{-1,1\}$. The Haldane-Shastry ground state can then be expressed as
\begin{equation}
|\psi_{0} \rangle = \sum_{s_1,s_2,\ldots,s_N}
\psi_{0}(s_{1},s_{2},\ldots,s_{N})
|s_{1},s_{2},...,s_N\rangle,
\end{equation}
where
\begin{multline}\label{psiHS}
\psi_{0}(s_{1},s_{2},\ldots,s_{N})=\\
\delta_{s} \prod_{k=1}^N e^{i \pi (k-1)(s_{k}+1)/2} \prod_{i<j} (z_{i}-z_{j})^{(s_{i}s_{j}-1)/2}.
\end{multline}
In this expression, $z_j$ is the position of the $j$th spin, and $N$ must be even, as $\delta_{s}$ is defined such that
\begin{equation}
\delta_{s} =
\left\{ \begin{array}{ll}
    1 & \textrm{if } \sum_{i} s_{i}=0\\
    0 & \textrm{otherwise}\\
\end{array}\right..
\end{equation}
In the original Haldane-Shastry model, $z_{j}=e^{i 2\pi j/N}$, which means that the spins are uniformly spaced. We shall here consider the more general case, where $z_{j}=e^{i \phi_{j}}$ with $\phi_{j} \in [0,2\pi[$. In both cases, \eqref{psiHS} is a spin singlet.

We start from the uniform case and add disorder by choosing the lattice positions as
\begin{equation}\label{zdisorder}
z_{j}=e^{2\pi i(j+\alpha_{j})/N}, \qquad
j\in\{1,2,\ldots,N\}.
\end{equation}
Here, $\alpha_{j}$ is a random number chosen with constant probability density in the interval $[-\case{\delta}{2},\case{\delta}{2}]$, and $\delta\in [0,N]$ is the disorder strength. An example for $N=6$ spins is shown in the right panel of Fig.\ \ref{fig:circle}. For disorder strength $\delta=0$, the uniform model is recovered. When $0<\delta<1$, there is some disorder, but the spins are still ordered from $1$ to $N$ on the circle, and there is still one spin per $2\pi/N$ length of the circle. If we increase the disorder strength even further to $\delta>1$, the order of the spins can change. For the maximal disorder strength $\delta=N$, all of the spins can be anywhere on the circle.

As long as the $z_j$ are restricted to be on the unit circle, one can show analytically that $|\psi_{0}\rangle$ is a ground state of the Hamiltonian
\begin{equation}\label{ham}
H=-2 \sum_{i\neq j}\bigg[\dfrac{z_{i}z_{j}}{(z_{i}-z_{j})^{2}} +  \dfrac{(z_{i}+z_{j})(c_{i}-c_{j})}{12(z_{i}-z_{j})}\bigg]\vec{S}_{i} \cdot \vec{S}_{j},
\end{equation}
and numerical computations show that it is the unique ground state \cite{nielsen2011quantum}. In this expression,
\begin{equation}
c_{j}=\sum_{k \in \{1,2,...,N\} \backslash \{j\}} \dfrac{z_{k}+z_{j}}{z_{k}-z_{j}}
\end{equation}
and $\vec{S}_{i}$ is the spin operator $(S_{i}^{x},S_{i}^{y},S_{i}^{z})$. Each $S_{i}^{a}$ is defined as
\begin{equation*}
S_{i}^{a} = \frac{1}{2} \sigma_i^a, \quad a\in\{x,y,z\},
\end{equation*}
where $\sigma^{x}, \sigma^{y}, \sigma^{z}$ are the Pauli matrices. We write the Hamiltonian in \eqref{ham} as
\begin{equation}\label{Hamilton}
H= \sum_{i\neq j}h_{ij}\vec{S}_{i} \cdot \vec{S}_{j}.
\end{equation}
The Hamiltonian conserves the net magnetization and we shall assume throughout that the net magnetization is fixed to zero.

\begin{figure}
\includegraphics[width=\columnwidth]{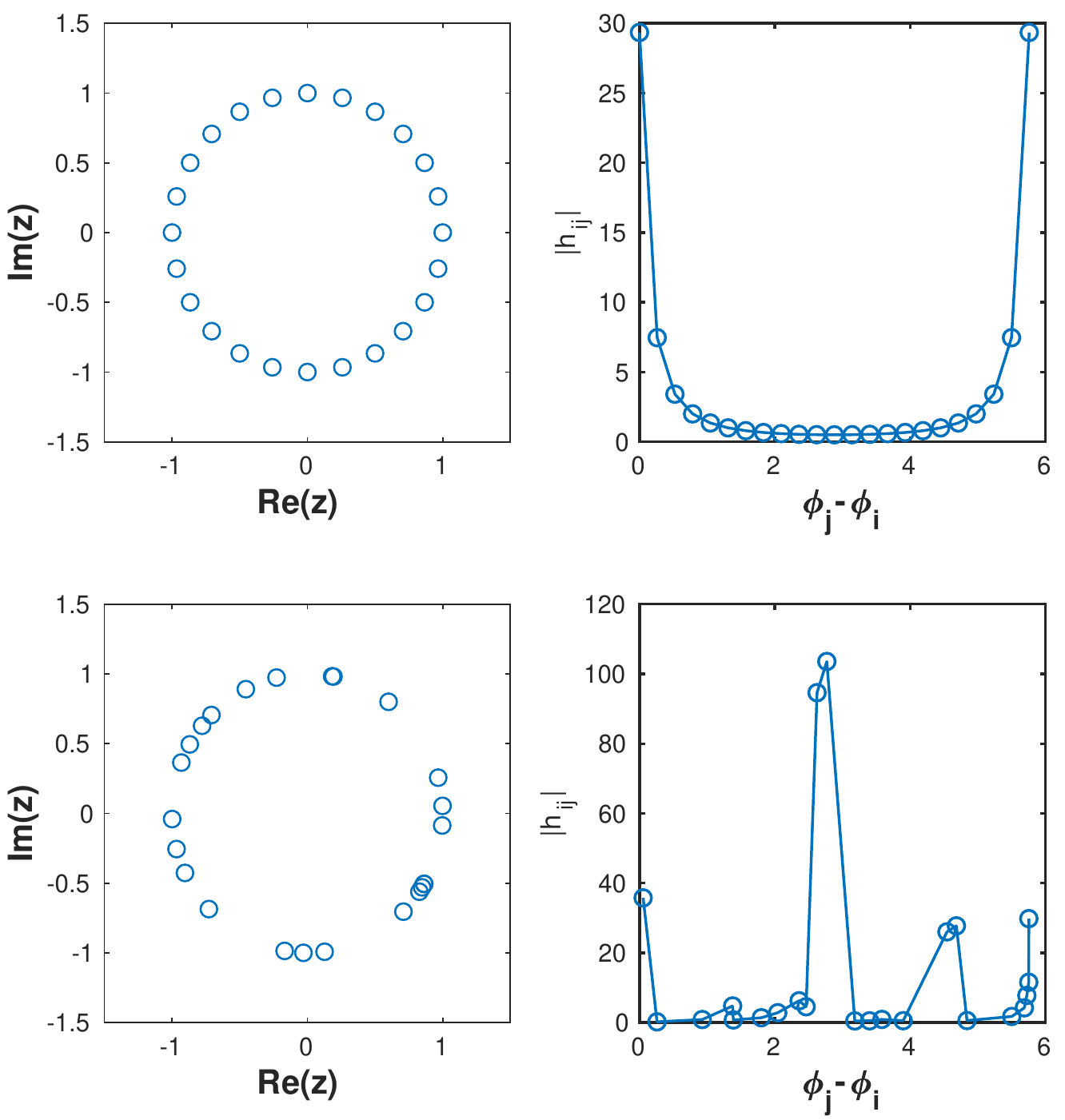}
\caption{Top left: Uniform lattice with $N=20$ spins. Top right: Behavior of the coupling strengths $|h_{ij}|$ in the Hamiltonian on the clean lattice. Bottom left: One realization of the disordered lattice with $N=20$ spins for maximum disorder strength $\delta=N$. Bottom right: Behavior of the coupling strengths for the disordered lattice shown on the left. In both the clean and disordered cases, we fix the $i^{\textrm{th}}$ spin to be the one with the smallest angle $\phi_i$ and plot $|h_{ij}|$ as a function of $\phi_j-\phi_i$. Note that for the disordered case, the coupling strength can be quite large between spins that are far from each other.}
\label{fig:coeff}
\end{figure}

We add disorder to this Hamiltonian in the same way as to the Haldane-Shastry ground state, namely by defining the lattice positions $z_j$ as in Eq.\ \eqref{zdisorder}. It is seen that disordering the positions of the spins leads to a disordering of the coupling strengths. The coupling strengths can be arbitrarily large when the spins come close to each other. Fig.\ \ref{fig:coeff} shows the behavior of the coupling coefficients with distance for cases with and without disorder. Although the coupling coefficients quickly decay to zero for the clean case, this is no longer true at strong disorder as seen in the figure.

\section{Properties of the disordered model}\label{sec:exc}

In the following, we study the properties of the highly excited states. The Hamiltonian can be truly long-ranged at strong disorder, and we find that the system is non-ergodic, but not many-body localized. This is similar to the findings in short-ranged SU(2) invariant models discussed in \cite{SU2,nonab}. In particular, we investigate numerically the entanglement entropy and the level spacing statistics.

\begin{figure}
\includegraphics[width=\columnwidth]{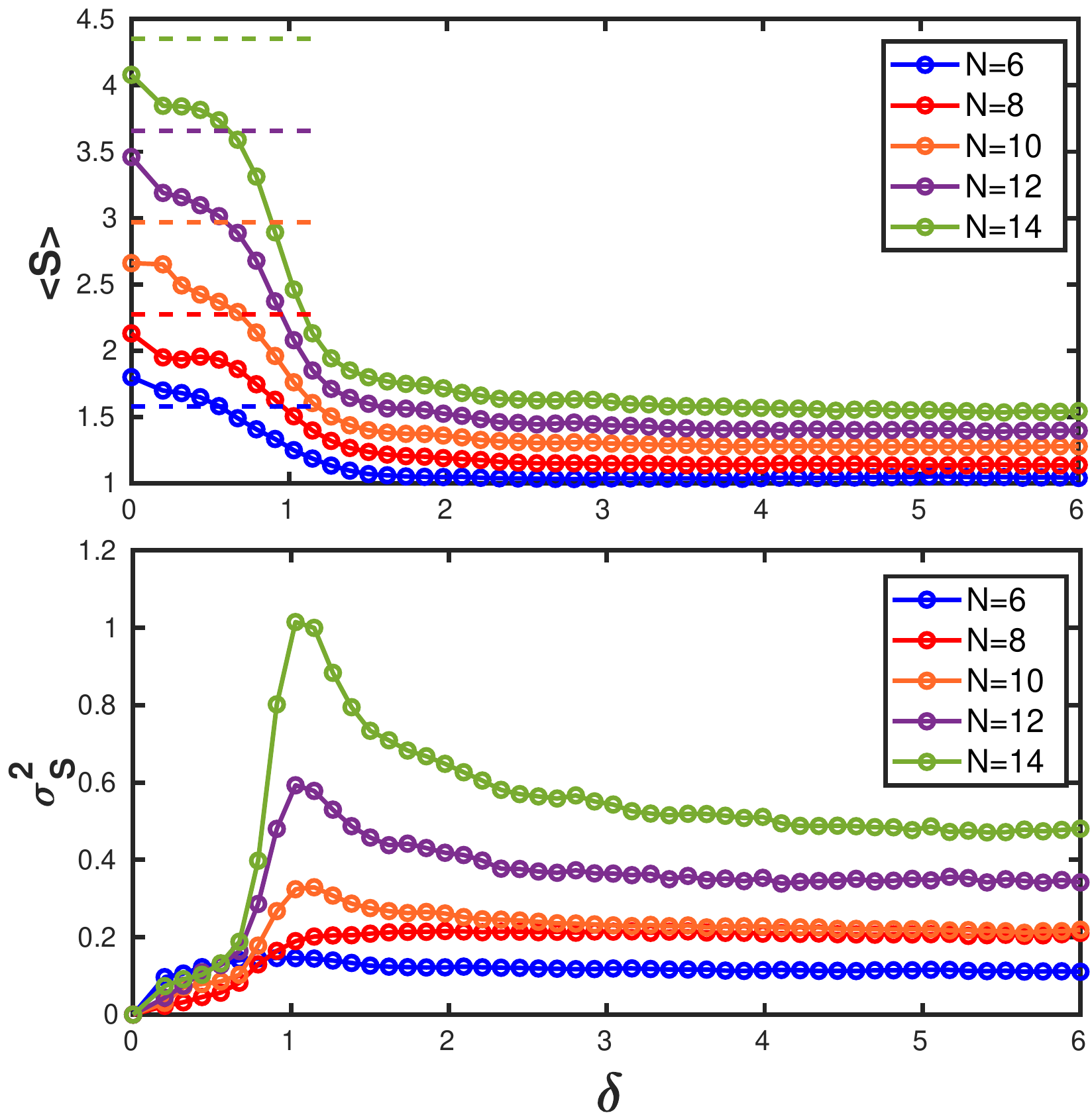}
\caption{Top: Entanglement entropy $S$ of half of the chain for the state closest to the middle of the spectrum of the Hamiltonian \eqref{ham} averaged over $10^4$ disorder realizations as a function of the disorder strength $\delta$. The different curves are for different system sizes $N$ as indicated. The entropy $[N \ln(2)-1]/2$ for an ergodic system is shown with horizontal dashed lines. Bottom: Variance $\sigma^2$ of the entanglement entropy computed from the same set of data. It is seen that the transition happens around $\delta=1$.} \label{fig:meanalt}
\end{figure}

\subsection{Entanglement entropy} \label{sec:transition}

The entanglement entropy of excited states has turned out to be a helpful diagnostics to identify phase transitions in disordered systems \cite{Bard}. Here, we consider the half chain entanglement entropy $S = -\mathrm{Tr}[\rho\ln(\rho)]$, where $\rho=\mathrm{Tr}_{B}(|\psi \rangle \langle \psi|)$ is the reduced density matrix after tracing out part $B$ of the system and $|\psi\rangle$ is the considered energy eigenstate. The manner in which we choose the spins constituting part $B$ of the chain is explained in Fig.\ \ref{fig:circle}. For each disorder realization, we pick the eigenstate whose energy is closest to the middle of the spectrum, i.e.\ closest to the energy $(E_\textrm{max} + E_\textrm{min})/2$, where $E_{\textrm{min}}$ is the ground state energy and $E_{\textrm{max}}$ is the highest energy in the spectrum.

\begin{figure}
\includegraphics[width=\columnwidth]{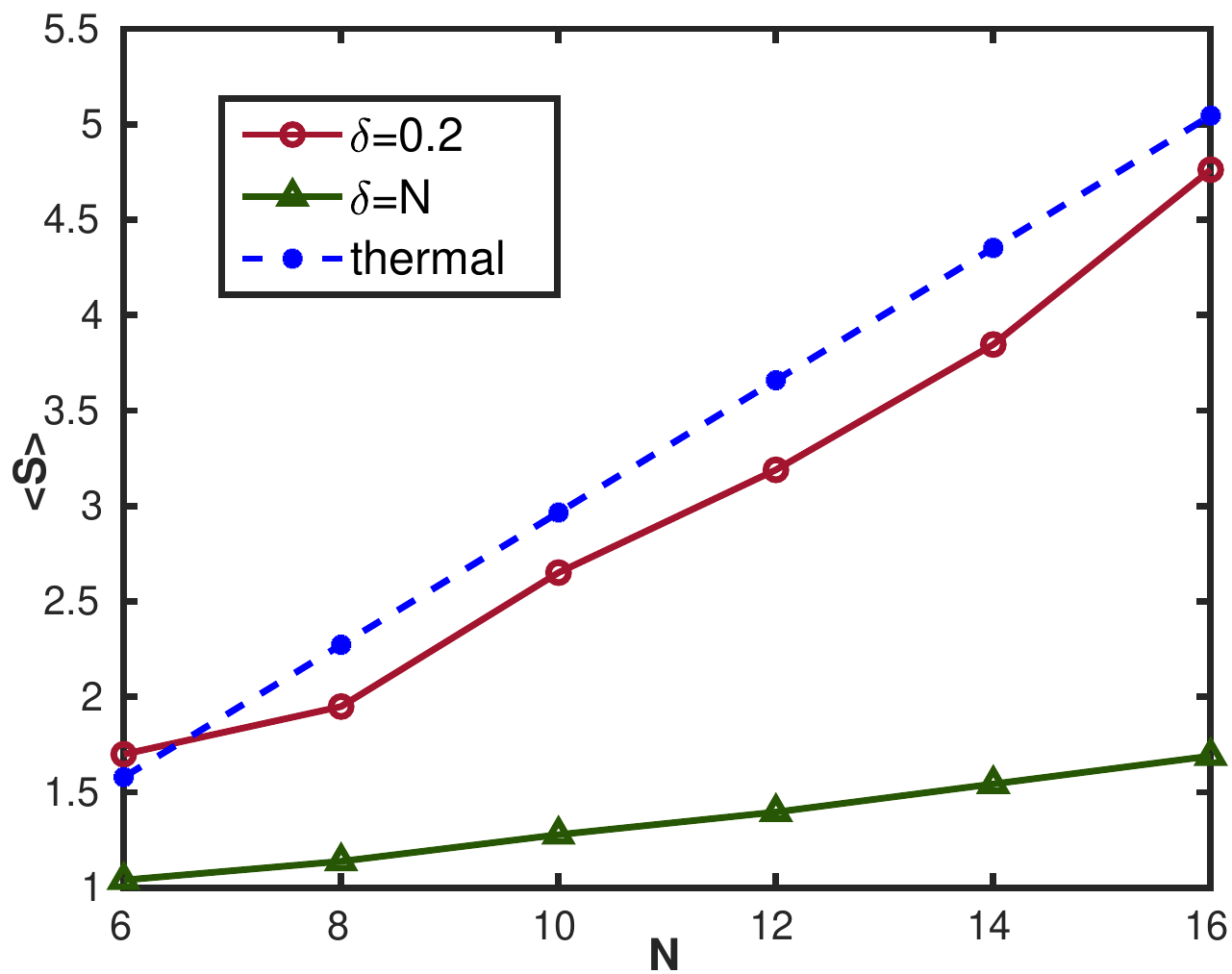}
\caption{Scaling of the disorder averaged half chain entanglement entropy $\langle S \rangle$ with the system size $N$ for weak and strong disorder. The slope of a linear fit for the points computed at weak disorder is around 0.3, while for those computed at maximum disorder it is around 0.06. The dashed blue line shows the entropy $[N \ln(2)-1]/2$ of an ergodic system for comparison. The results shown are averaged over $10^4$ disorder realizations.}\label{fig:entscale}
\end{figure}

In Fig.\ \ref{fig:meanalt}, we plot the mean entanglement entropy and the variance of the distribution as a function of the disorder strength $\delta$ for different system sizes. The plots show a phase transition at $\delta \approx 1$. The reason why the phase transition happens around that point could be due to the fact that for $\delta\geq1$, neighboring spins can be arbitrarily close, while this is not the case for $\delta<1$.

Figure \ref{fig:entscale} shows how $\langle S \rangle$ scales with the system size $N$ on the weak and strong disorder side of the transition and compares these results to the entropy of an ergodic system of $N$ spin-$1/2$, which was conjectured to be $[N \ln(2)-1]/2$ for $2^{N/2}\gg 1$ in \cite{page}. At small disorder, the entropy grows with the system size, and the slope is comparable to the slope for the entropy of an ergodic system. At large disorder, the entropy again increases linearly with the system size, which shows that the system is not many-body localized. The slope is, however, much less than it is for weak disorder, and the system is therefore also not ergodic.

\begin{figure}
\includegraphics[width=\columnwidth]{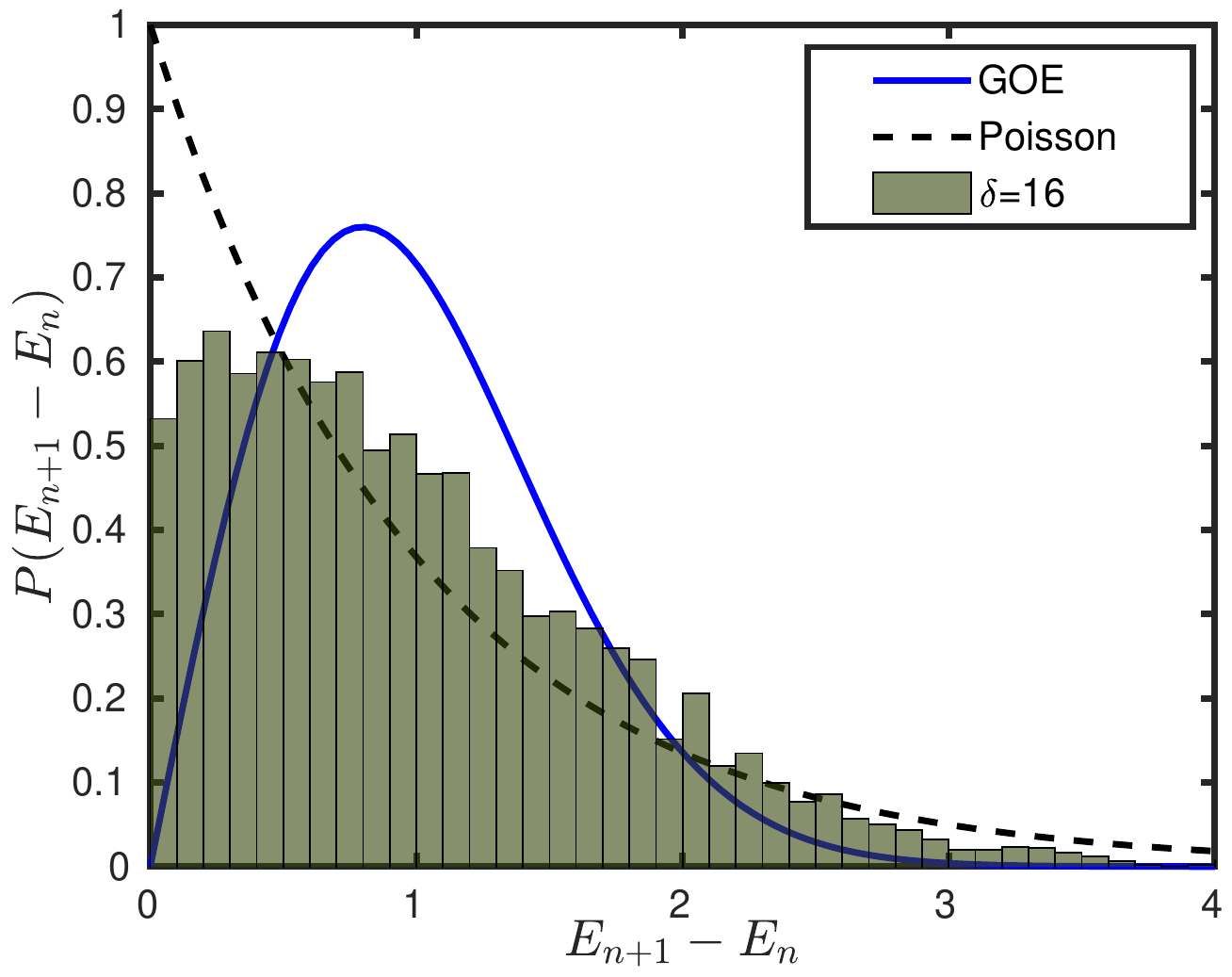}
\caption{The energy level spacing distribution of the disordered Haldane-Shastry model computed for 16 spins for one disorder realization at maximal disorder strength. It follows neither the Poisson distribution, nor the Gaussian orthogonal ensemble.}
\label{fig:dist}
\end{figure}

\subsection{Level spacing statistics}

\begin{figure}
\includegraphics[width=\columnwidth]{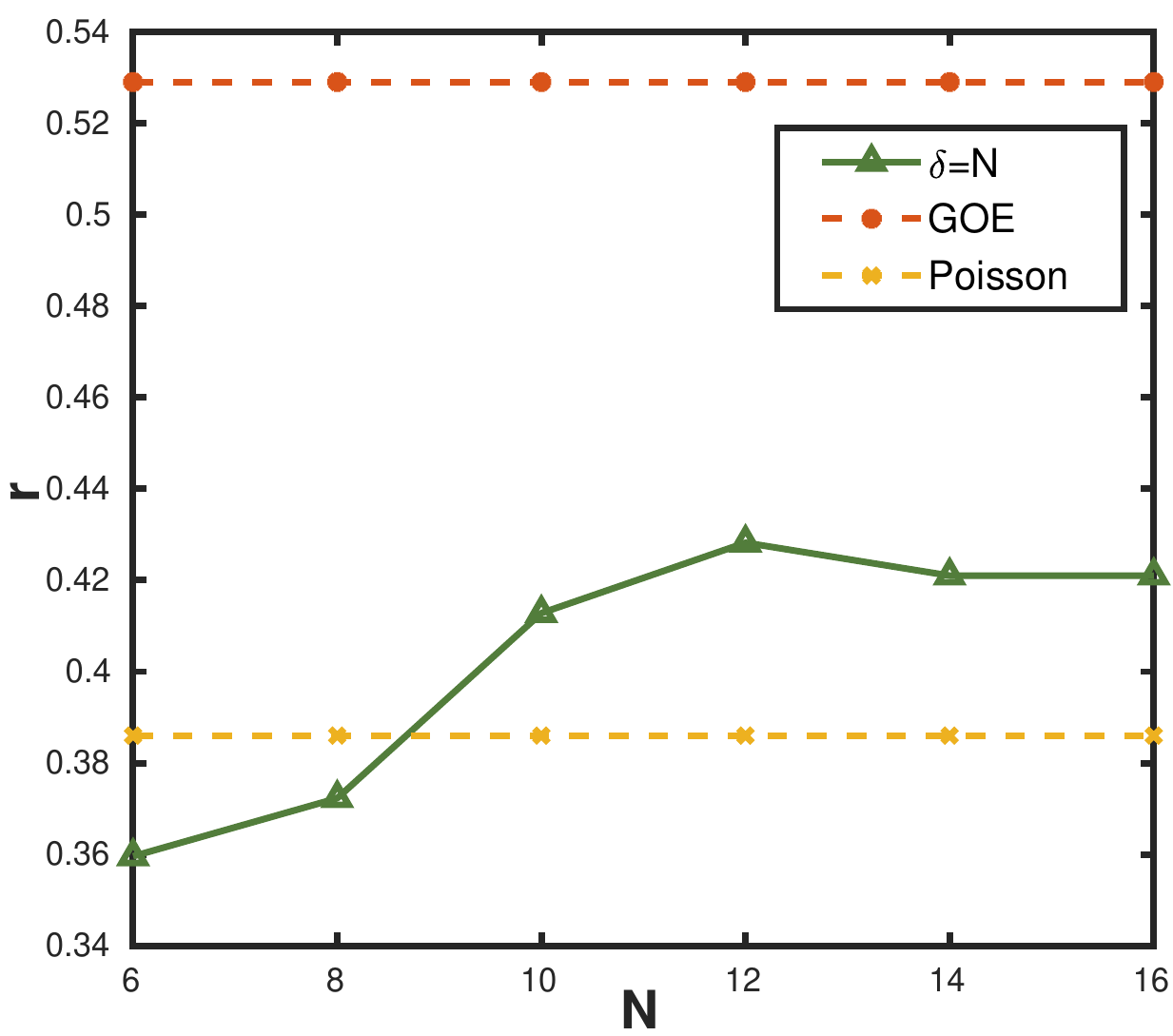}
\caption{The adjacent gap ratio $r$ for the spectrum of the Hamiltonian \eqref{ham} as a function of the system size $N$ at maximal disorder strength. The results shown are averaged over $10^4$ disorder realizations. The curve neither approaches the Poisson value, nor the value for the Gaussian orthogonal ensemble.}
\label{fig:agralt}
\end{figure}

The level spacing statistics is another helpful tool to identify whether a system is in an ergodic phase, a many-body localized phase, or in some other phase. Hamiltonians with or without time reversal symmetry \cite{Nand} in the ergodic phase are known to exhibit Gaussian orthogonal ensemble or Gaussian unitary ensemble level spacing statistics, respectively, indicating the presence of energy level repulsion. While in the localised phase, the level spacing statistics obey the Poisson distribution indicating the absence of level repulsion. The energy level spacing distribution of the clean Haldane-Shastry model has been studied in the past and is found not to obey the Poissonian distribution although the model is completely integrable \cite{HSls,Hsu}. We compute the distribution $P(E_{n+1}-E_n)$ of the energy level spacings $E_{n+1}-E_n$ of the disordered Haldane-Shastry model for the maximally disordered case and find that, interestingly, it is neither of the Wigner-Dyson, nor of the Poisson kind, which is seen in Fig.\ \ref{fig:dist}.

One can also compute a dimensionless quantity, which is the ratio of consecutive gaps of distinct energy levels \cite{Huse}. The quantity of interest is called the adjacent gap ratio and is defined as follows
\begin{equation}\label{gapratio}
r=\frac{1}{N_s}\sum_{n} \frac{\min(\delta_{n},\delta_{n-1})}{\max(\delta_{n},\delta_{n-1})},
\end{equation}
where $\delta_{n}=E_{n+1}-E_{n}$ and $N_s$ is the total number of states in the spectrum. The value of $r$ averaged over several disorder realizations is around $0.529$ for the Gaussian orthogonal ensemble and around $0.386$ for the Poisson distribution. We recall that we are considering the sector of the Hilbert space with zero net magnetization. Also, since the Hamiltonian commutes with the parity operator $\prod_i\sigma^i_x$ , we consider the energies in one of the parity sectors for computing the level spacing statistics. The plot in Fig.\ \ref{fig:agralt} shows the gap ratio computed for several system sizes at maximum disorder strength. The gap ratio is not close to either $0.529$ or $0.386$. This indicates that the system is in a non-ergodic phase that does not exhibit many-body localization. This is consistent with the sub-thermal value of the half-chain entanglement entropy at strong disorder observed above.

\begin{figure}
\includegraphics[width=\columnwidth]{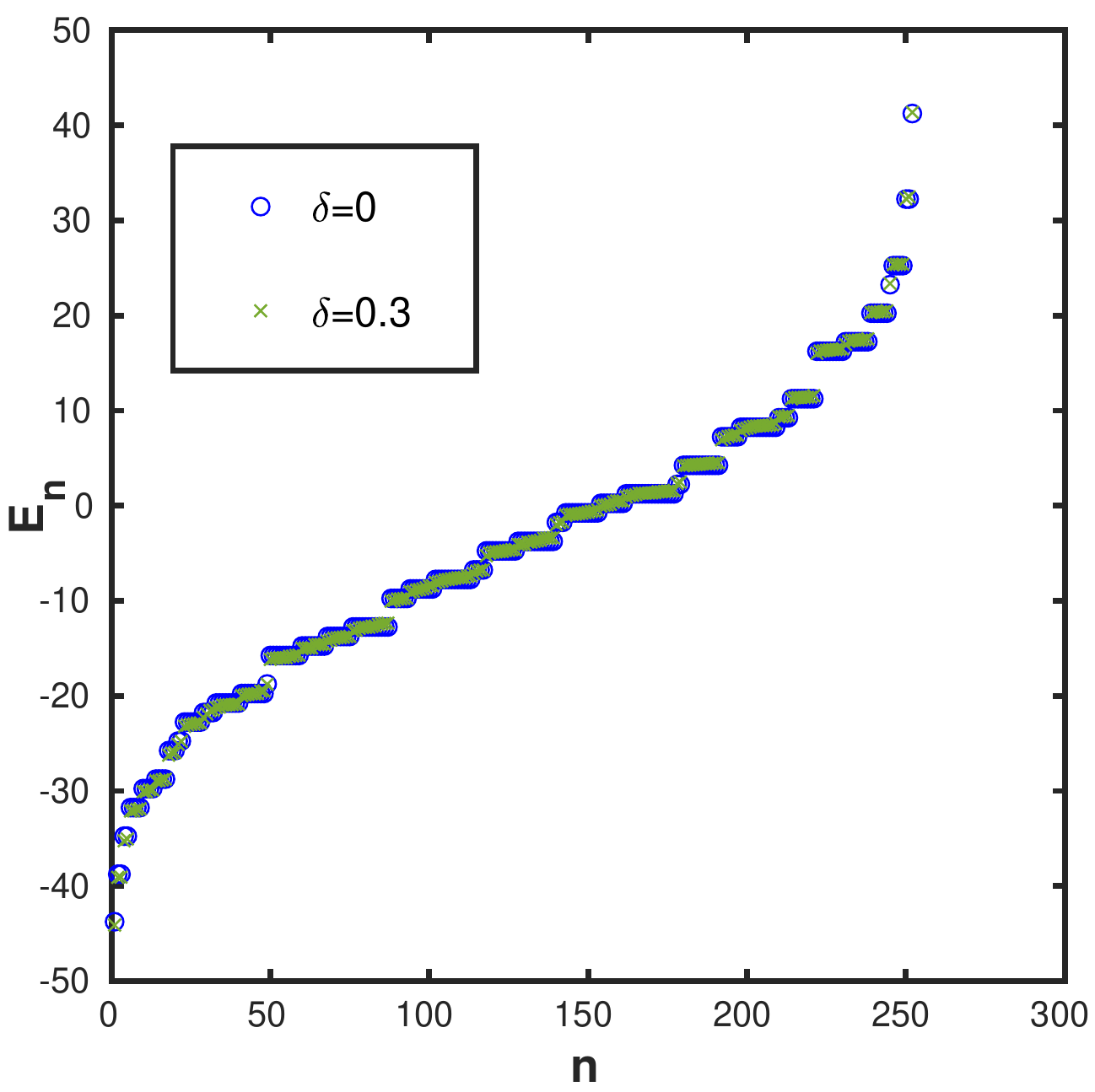}
\caption{Energy spectrum of the Haldane-Shastry model at no and weak disorder computed for 12 spins. The spectrum at weak disorder closely resembles that of the clean Haldane-Shastry model.}\label{fig:spect}
\end{figure}

We avoid showing the level spacing statistics for weak disorder for the following reason. The clean Haldane-Shastry model has an enhanced Yangian symmetry generated by the total spin operator and the rapidity operator that both commute with the Hamiltonian \cite{talstra}, and this leads to degeneracies in the spectrum. For weak disorder, the model has no exact Yangian symmetry, but there are still approximate degeneracies visible in the energy spectrum. This is seen in Fig.\ \ref{fig:spect}, where we compare the energy spectra of the clean model and the disordered model at weak disorder. The energy spectrum with little disorder closely resembles the spectrum of the clean Haldane-Shastry model. The degeneracies, although not exact, have an impact on the level spacing statistics and make this measure more difficult to interpret.

\section{Adding disorder to the excited states of the Haldane-Shastry model}\label{sec:last}

All the excited states of the clean model can be analytically expressed in terms of the positions $z_j$ \cite{Herwerth}. This opens up the possibility to add disorder into the excited states by modifying $z_j$. It turns out, however, that the resulting states $|\psi'_n\rangle$ are not eigenstates of the Hamiltonian \eqref{ham} as soon as $\delta>0$. If the states had been orthonormal, one could construct an alternative Hamiltonian as $H=\sum_n E_n |\psi'_n\rangle\langle\psi'_n|$, where $E_n$ is the energy of the state $|\psi'_n\rangle$. We find numerically, however, that most of the states are not orthogonal for $\delta>0$. The most appropriate way to add disorder into the model is hence to introduce it in the Hamiltonian \eqref{ham} and study the properties of the eigenstates numerically as discussed in the previous sections.

\section{Conclusion}
\label{sec:conclusion}

We investigated the role of disorder on the Haldane-Shastry model, which has long-range interactions and SU(2) symmetry. The disorder averaged entanglement entropy for a state close to the middle of the energy spectrum showed a transition into a non-ergodic phase at strong disorder. The non-ergodic phase is not a many-body localized phase, since it has volume law entanglement. The coefficient of the volume law is, however, much less than for an ergodic system. We further confirmed this non-ergodic behavior from the level spacing statistics that did not match the random matrix predictions for a thermal or a many-body localized phase.

The behaviors of strongly-correlated quantum many-body systems are affected by several factors, including symmetries and the range of the terms in the Hamiltonian. It has been argued that SU(2) symmetry is not compatible with many-body localization \cite{potter,vasseur}, and it is hence expected that the disordered Haldane-Shastry model does not many-body localize. On the other hand, low entanglement and non-ergodic behaviors, which are different from many-body localization, have been found in a nearest neighbor model with SU(2) symmetry \cite{SU2,nonab}. There have been several studies investigating whether many-body localization can survive in the presence of different types of long-range terms in the Hamiltonian \cite{yao,singh,Nand,nag}, since one might expect that long-range terms will increase the amount of entanglement. Similarly, it is unclear, whether nonlocal terms will destroy the non-ergodic, low entanglement behavior seen in local models with SU(2) symmetry. The results presented here provide an example, which shows that the non-ergodic, low entanglement behavior can also happen in nonlocal models. The obtained results motivate the search for further, interesting, non-ergodic phases in models with SU(2) symmetry and long-range interactions, as well as systematic studies to clarify under which types of conditions, models with non-ergodic behavior may appear.

%

\end{document}